\documentclass[%
 reprint,
 amsmath,amssymb,superscriptaddress,
 aps,prl,
float,
]{revtex4-1}
\usepackage{color}
\usepackage{graphicx}
\usepackage{dcolumn}
\usepackage{bm}

\begin{document}

\title{Anomalous helimagnetic domain shrinkage due to the weakening of Dzyaloshinskii-Moriya interaction in CrAs}

\author{B. Y. Pan}\thanks{B.P. and H.X. contributed equally to this work. D.F. conceived the idea and supervised the project.}
\affiliation{State Key Laboratory of Surface Physics, Department of Physics, and Advanced Materials
Laboratory, Fudan University, Shanghai 200433, China}
\affiliation{School of Physics and Optoelectronic Engineering, Ludong University, Yantai, Shandong 264025, China}
\author{H. C. Xu}\thanks{B.P. and H.X. contributed equally to this work. D.F. conceived the idea and supervised the project.}
\affiliation{State Key Laboratory of Surface Physics, Department of Physics, and Advanced Materials
Laboratory, Fudan University, Shanghai 200433, China}
\author{Y. Liu}
\affiliation{Center for Correlated Matter, Zhejiang University, Hangzhou, 310058, China}
\author{R. Sutarto}
\author{F. He}
\affiliation{Canadian Light Source, Saskatoon, Saskatchewan S7N 2V3, Canada}
\author{Y. Shen}
\affiliation{State Key Laboratory of Surface Physics, Department of Physics, and Advanced Materials
Laboratory, Fudan University, Shanghai 200433, China}
\author{Y. Q. Hao}
\affiliation{State Key Laboratory of Surface Physics, Department of Physics, and Advanced Materials
Laboratory, Fudan University, Shanghai 200433, China}
\author{J. Zhao}
\affiliation{State Key Laboratory of Surface Physics, Department of Physics, and Advanced Materials
Laboratory, Fudan University, Shanghai 200433, China}
\author{Leland Harriger}
\affiliation{NIST Center for Neutron Research, National Institute of Standards and Technology, 100 Bureau Drive, Gaithersburg, Maryland 20899, USA}
\author{D. L. Feng}\email{dlfeng@fudan.edu.cn}
\affiliation{State Key Laboratory of Surface Physics, Department of Physics, and Advanced Materials
Laboratory, Fudan University, Shanghai 200433, China}
\affiliation{Collaborative Innovation Center of Advanced Microstructures, Nanjing 210093, China}
\affiliation{Hefei National Laboratory for Physical Science at Microscale and Department of Physics, University of Science and Technology of China, Hefei, Anhui 230026, China}

\begin{abstract}

CrAs is a well-known helimagnet with the double-helix structure originating from the competition between the Dzyaloshinskii-Moriya interaction (DMI) and antiferromagnetic exchange interaction $J$. By resonant soft X-ray scattering (RSXS), we observe the magnetic peak (0~0~$q_m$) that emerges at the helical transition with $T_S$ $\approx$ 267.5 K. Intriguingly, the helimagnetic domains significantly shrink on cooling below $\sim$255~K, opposite to the conventional thermal effect. The weakening of DMI on cooling is found to play a critical role here. It causes the helical wave vector to vary, ordered spins to rotate, and extra helimagnetic domain boundaries to form at local defects, thus leading to the anomalous shrinkage of helimagnetic domains. Our results indicate that the size of magnetic helical domains can be controlled by tuning DMI in certain helimagnets.
\end{abstract}
\maketitle
In correlated materials, multiple magnetic interactions, including the superexchange, Dzyaloshinskii-Moriya interaction (DMI),  Kondo coupling, Ruderman$-$Kittel$-$Kasuya$-$Yosida interaction, may co-exist, and they favor different ground states. The competition between these magnetic interactions leads to rich and novel phenomena such as the quantum criticality in Kondo lattice\cite{RevModPhys.79.1015}, spin liquid states in frustrated systems\cite{Broholmeaay0668}, and the emergence of magnetic skyrmions in helimagnets\cite{Nature442,PhysRevB.82.094429}. Tuning the strength of these interactions would be a important way to engineer the magnetic quantum states and properties. Take a helimagnet system for example, its magnetic Hamitonian can be generally written as:
\begin{equation}
 H=\sum_{i,j}\vec{D}_{ij}\cdot(\vec{S}_i\times\vec{S}_j)+J_{i,j}\vec{S}_i\cdot\vec{S}_j
\end{equation}
in which $\vec{D}_{i,j}$ and $J_{i,j}$ denote the anti-symmetric DMI and the symmetric exchange interactions betweeen $\vec{S}_i$ and $\vec{S}_j$, respectively. By changing temperature, magnetic field, material thickness, or pressure, the system can be continuously tuned into helical, conical, Skyrme
crystal or other quantum phases depending on the subtle balance of DMI, $J$, Zeeman coupling, and thermal fluctuations\cite{Motizuki2010,Sci2009,PhysRevB.96.214425,PhysRevLett.115.267202}. Especially, the nano-sized helimagnetic domains, a key ingredient for spintronics \cite{Schoenherr2018}, can be delicately manipulated by external fields. For example, in-situ Lorentz microscopy of Fe$_{0.5}$Co$_{0.5}$Si film showed that magnetic field can effectively deform, rotate and enlarge the helimagnetic domains by applying $H$ along different directions\cite{Uchida2006}. Similar observation was recently reported in Te-doped Cu$_2$OSeO$_3$\cite{Haneaax2138}. Here, in the helimagnet CrAs, we observe an anomalous shrinkage of helimagnetic domains with decreasing temperature and find the decrease of DMI in CrAs on cooling as its main driving force. This is a quantum effect opposite to conventional thermal behavior and may be harnessed for domain engineering in spintronics.

The helical transition temperature of CrAs is $T_S$ = 265 K and its spin helix propagates along the $c$ axis (Fig. 1(b))\cite{Boller1971,Motizuki2010,Shen2016}. The local space-inversion symmetry breaking at the Cr and As sites gives rise to DMI, which is essential in stabilizing the doule-helix spin structure as revealed by group-theoretical approach\cite{Kallel1974}. Compared with other helimagnets in the MnP-type structure, such as MnP and FeP, the DMI in CrAs is much more pronounced\cite{Kallel1974,Sjostrom1992}. In addition, recent studies show CrAs exhibits novel non-Fermi liquid behavior, unconventional superconductivity, and quantum criticality under certain conditions \cite{Wu2014,Shen2016,Matsuda2018,Kotegawa2015,Guo2018,Park2019}, and the helical magnetism is believed to be crucial on these fascinating properties\cite{Norman2015}. The strong DMI and rich quantum phases in CrAs make it an exciting playground to study the behavior of magnetic domains under the competition between DMI and $J$.

In our experiment, we used soft X-ray absorption and resonant scattering to study the helical magnetism of CrAs. The magnetic resonant peak (0~0~$q_m$) was observed at the chromium $L$-edge. Thanks to the high momentum resolution of resonant soft X-ray scattering (RSXS) technique, we could reveal the average helimagnetic domain size $\xi$ and its temperature-dependent evolution. Intriguingly, unlike conventional magnets whose magnetic domains always grow larger on cooling, the domain size of CrAs substantially $decreases$ with lowering temperature below $\sim$255 K. The average domain size in the $ac$ plane shrinks by $\sim$ 44.14\% from 255 K to 20 K. We find that the temperature dependent domain shrinkage follows the weakening of DMI, whose competition with $J$ varies the helimagnetic wave vector. As the helical magnetic chains propagate across the crystal defects with decreasing DMI on cooling, extra helix domain boundaries form and the average domain size decreases, leading to the observed domain shrinkage opposite to the conventional thermal effect.

\begin{figure}[]
\centering
\includegraphics[clip,width=8cm]{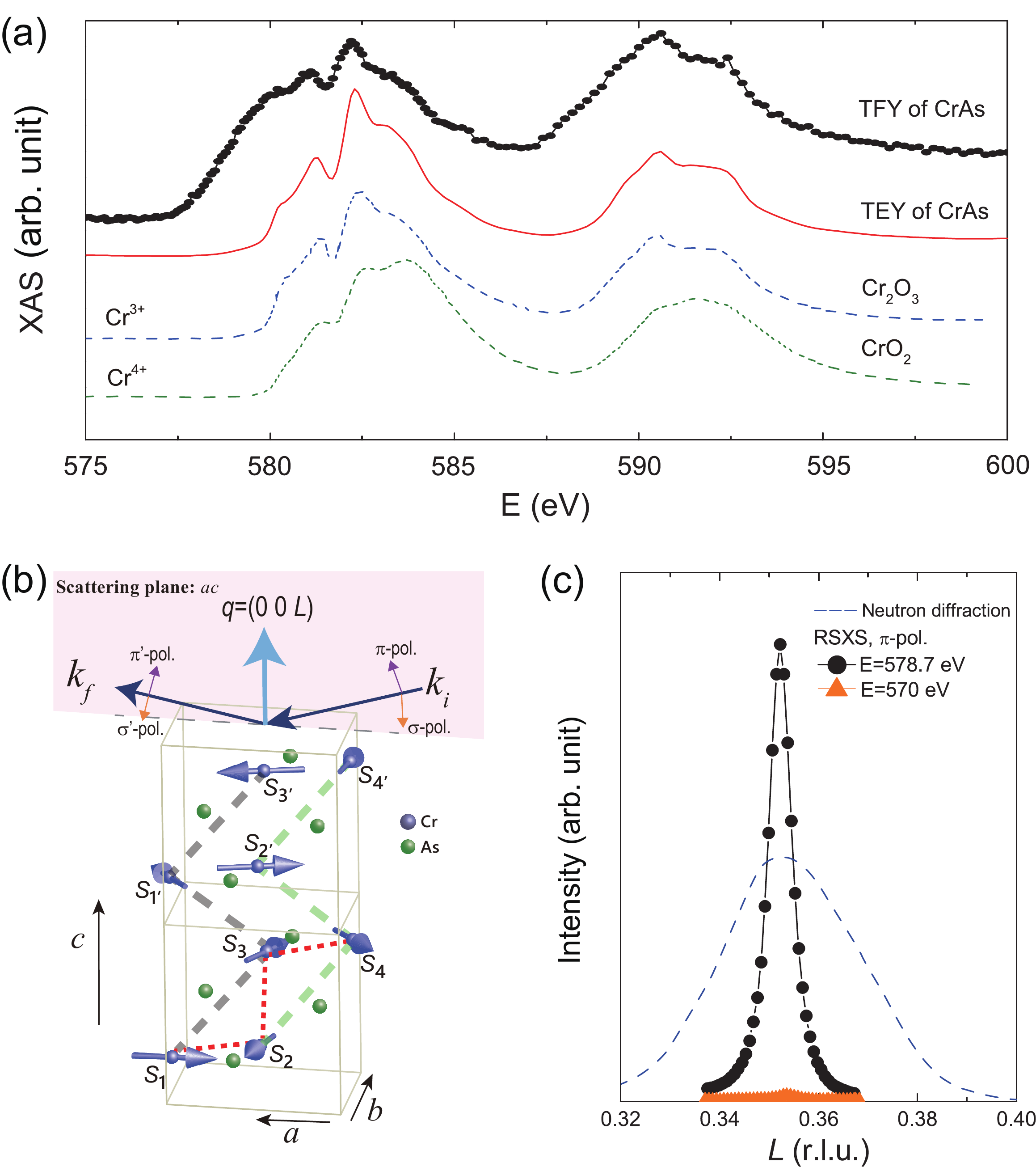}
\caption{\label{fig:epsart} (a) X-ray absorption spectroscopy (XAS) in total electron yield (TEY) mode of CrAs at the Cr $L$-edge (red solid line), in comparison with two reference compounds Cr$_2$O$_3$ (blue dashed line) and CrO$_2$ (Olive dashed line) from literature\cite{Dedkov2005}. The bulk sensitive total fluorescence yield  (TFY) spectra of CrAs (black dots) was simultaneously collected. (b) Illustration of the helical spin order in CrAs and the RSXS scattering geometry. The gray and green dashed lines denote the double spin helix chains running along the $c$ axis. The red dashed lines denote the nearest-neighboring spins in a unit cell. (c) $L$ scans of the (0~0~$q_m$) magnetic peak with the resonant energy E=578.7 eV (black dots) and non-resonant energy E=570~eV (orange filled triangles) at $T$ = 20~K. The incident photons are $\pi$-polarized. The blue dashed line is from neutron diffraction measurement on a CrAs single crystal, note its position has been shifted by +0.007 r.l.u. in order to match the RSXS data.}
\end{figure}

CrAs single crystals were grown by the Sn-flux method described in previous report\cite{Wu2010}. The obtained shiny crystals have needle-like shape with a typical size of 6$\times$1$\times$0.5 mm$^3$. The largest crystalline plane is (0~0~1). RSXS and X-ray absorption (XAS) experiments on a CrAs single crystal were performed using a four-circle diffractometer at the Resonant Elastic and Inelastic X-ray Scattering (REIXS) beamline of Canadian Light Source (CLS). The REIXS beamline is equipped with Elliptical Polarized Undulators (EPU) and can provide both $\sigma$ and $\pi$ polarized incident photons. The momentum resolution of the RSXS instrument is better than 0.0005 \AA$^{-1}$ at 570 eV. For RSXS signal, a silicon photodiode was used, while for the XAS in total electron yield (TEY) mode and total fluorescence yield (TFY) mode were collected using a drain current and micro-channel plate (MCP) detector, respectively.

\begin{figure*}[]
\centering
\includegraphics[clip,width=18cm]{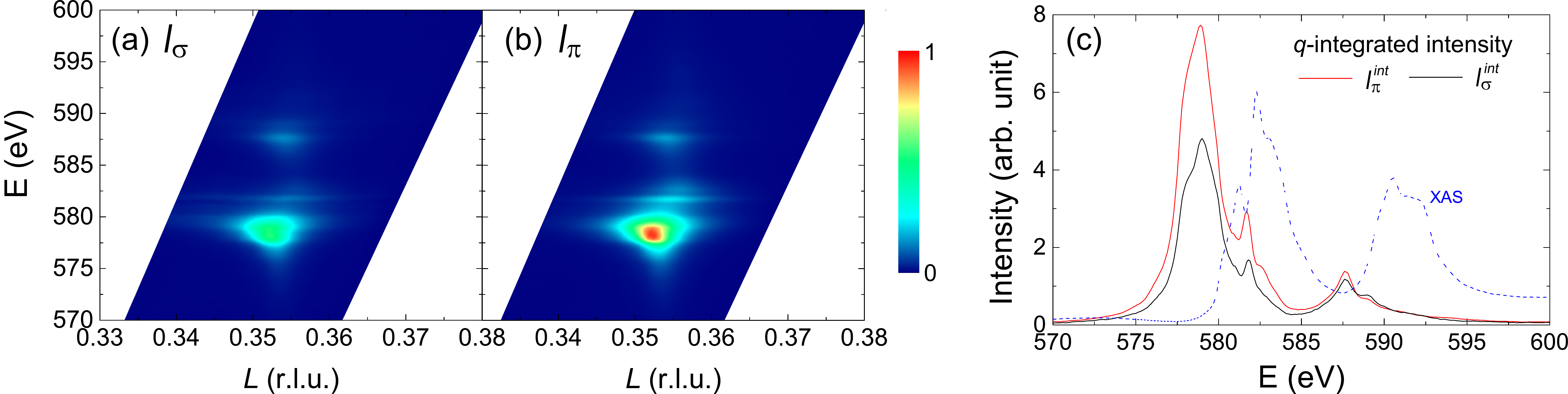}
\caption{\label{fig:epsart} Resonant profiles around (0 0 $q_m$) with (a) $\sigma$ or (b) $\pi$ polarized incident photons at $T$ = 20 K. The resonance profiles are plotted as a function of photon energy covering the Cr $L$-edge and reciprocal lattice (0~0~$L$). All data were measured by a photodiode detector and the fluorescence background has been subtracted. The color bar indicates scattering intensity in arbitrary unit. (c) Integrated intensity along the $L$ direction of the resonances in (a,b) as a function of the incident photon energy. The black and red lines represent the $\sigma$ and $\pi$ polarized incident photons, respectively. The dashed blue line is the XAS curve.}
\end{figure*}

Figure 1(a) is the XAS of CrAs (red line) measured by TEY at the Cr $L$-edge. In order to elucidate the Cr valance state, the XAS of Cr$_2$O$_3$ and CrO$_2$ were plotted together as the fingerprint of Cr$^{3+}$ and Cr$^{4+}$ valence states\cite{Dedkov2005}, respectively. The bulk sensitive TFY spectrum of CrAs was simultaneously collected. Both the TEY and TFY spectra of CrAs are consistent with the typical Cr$^{3+}$ spectrum. These results show that the sample surface is clean and the valance state is Cr$^{3+}$ with the 3d$^3$ electronic configuration. The spin-only magnetic moment of Cr$^{3+}$ is 3.87 $\mu_B$, however, the observed value is 1.7 $\mu_B$\cite{Selte1971}. The reduction of magnetic moment in CrAs should come from fluctuations and hybridization effect, similar to the case in MnP\cite{Pan2018}.

In our RSXS experiment, the scattering crystalline plane is $ac$ and the momentum transfer direction is along (0~0~$L$) (Fig. 1(b)). In this configuration, the electric field of horizontally (vertically) polarized incident photons are perpendicular (parallel) to the $b$ axis. In the helical state below $T_S$, the Cr$^{3+}$ spin moments lie in the $ab$ easy plane, and the magnetic propagation wavevector is about (0~0~0.356) at $T$=4 K\cite{Shen2016}. From sample alignment, we determined the lattice constants are $a$=5.412(9) \AA, $b$=3.348(1) \AA, $c$=6.007(9) \AA~at 20 K. Figure 1(c) presents the $L$ scans around the magnetic wavevector $k_m$=(0~0~$q_m$) with the resonant (E = 578.7 eV, $\pi$ polarization) and non-resonant (E = 570 eV, $\pi$ polarization) energies at $T$ = 20 K. Strong resonant peak appears around (0~0~0.352), consistent with the previously reported helical propagation wavevector\cite{Selte1971}. The blue dashed line in Fig. 1(c) is from neutron diffraction measurement on CrAs single crystal which is much broader due to its relatively low momentum resolution. A detailed comparison of neutron diffraction and RSXS experiments on CrAs can be found in the Supplemental Material.

To verify the magnetic nature of the (0~0~$q_m$) peak, we measured its resonant profiles as a function of X-ray energy and wavevector (0~0~$L$) at $T$ = 20 K (Fig. 2). The incident X-ray is either vertically ($\sigma$, Fig. 2(a)) or horizontally ($\pi$, Fig. 2(b)) polarized. No distinction was made on the polarization of scattered photons, so the detected scattering intensities in our experiment are $I_{\pi}$=$I_{\pi\pi'}$+$I_{\pi\sigma'}$ and $I_{\sigma}$=$I_{\sigma\pi'}$+$I_{\sigma\sigma'}$. The observed results show that $I_{\pi}$ is about 1.7 times stronger than $I_{\sigma}$. We further integrate the resonant intensity in Fig. 2(a) and 2(b) along the $L$ direction and get the $q$-integrated intensities $I_{\sigma}^{int}$ and $I_{\pi}^{int}$, as shown in Fig. 2(c). The lineshape profiles of $I_{\sigma}^{int}$ (black line) and $I_{\pi}^{int}$ (red line) are similar, except the latter is apparently stronger. The polarization dependence of  resonant profiles is consistent with magnetic scattering from a helimagnet, as evidenced by the following theoretical analysis. The scattering intensity from a helimagnet can be expressed as $I_{mag}$=$\vert$$f_{mag}$$\vert$$^2$, and $\vert$$f_{mag}$$\vert$ is the resonant magnetic scattering length\cite{Jang2016,Lovesey1996,Ramakrishnan2017}:
\begin{equation}
f_{mag}=\left(                 
  \begin{array}{cc}   
    f_{\sigma\sigma'} & f_{\pi\sigma'} \\  
    f_{\sigma\pi'} & f_{\pi\pi'}\\  
  \end{array}
\right)
\nonumber
\end{equation}

\begin{equation}       
\begin{split}
= -iF^1\left(                 
  \begin{array}{cc}   
    0 & M_acos\theta+M_csin\theta \\  
    M_csin\theta-M_acos\theta & -M_bsin2\theta\\  
  \end{array}
\right)             
\end{split}
\end{equation}
where $\sigma'$ and $\pi'$ denote the polarization of outgoing photons, $\theta$ is the angle between the incident X-ray and the sample surface, and $M_a$, $M_b$, $M_c$ are the spin moment along the three crystal axes. In our case, $M_a$=$M_b$, $M_c$=0, so $\vert$$f_{\sigma\pi'}$$\vert$=$\vert$$f_{\pi\sigma'}$$\vert$, $I_{\sigma}$=$I_{\sigma\pi'}$=$I_{\pi\sigma'}$. In this way, $I_{\pi}$=$I_{\sigma\pi'}$+$I_{\pi\pi'}$$>$$I_{\sigma}$, so $I_{\pi}$ is always stronger than $I_{\sigma}$ within the Cr $L$-edge resonant energy range, this is consistent with the experiment observation shown in Fig. 2.

\begin{figure}[]
\centering
\includegraphics[clip,width=8.5cm]{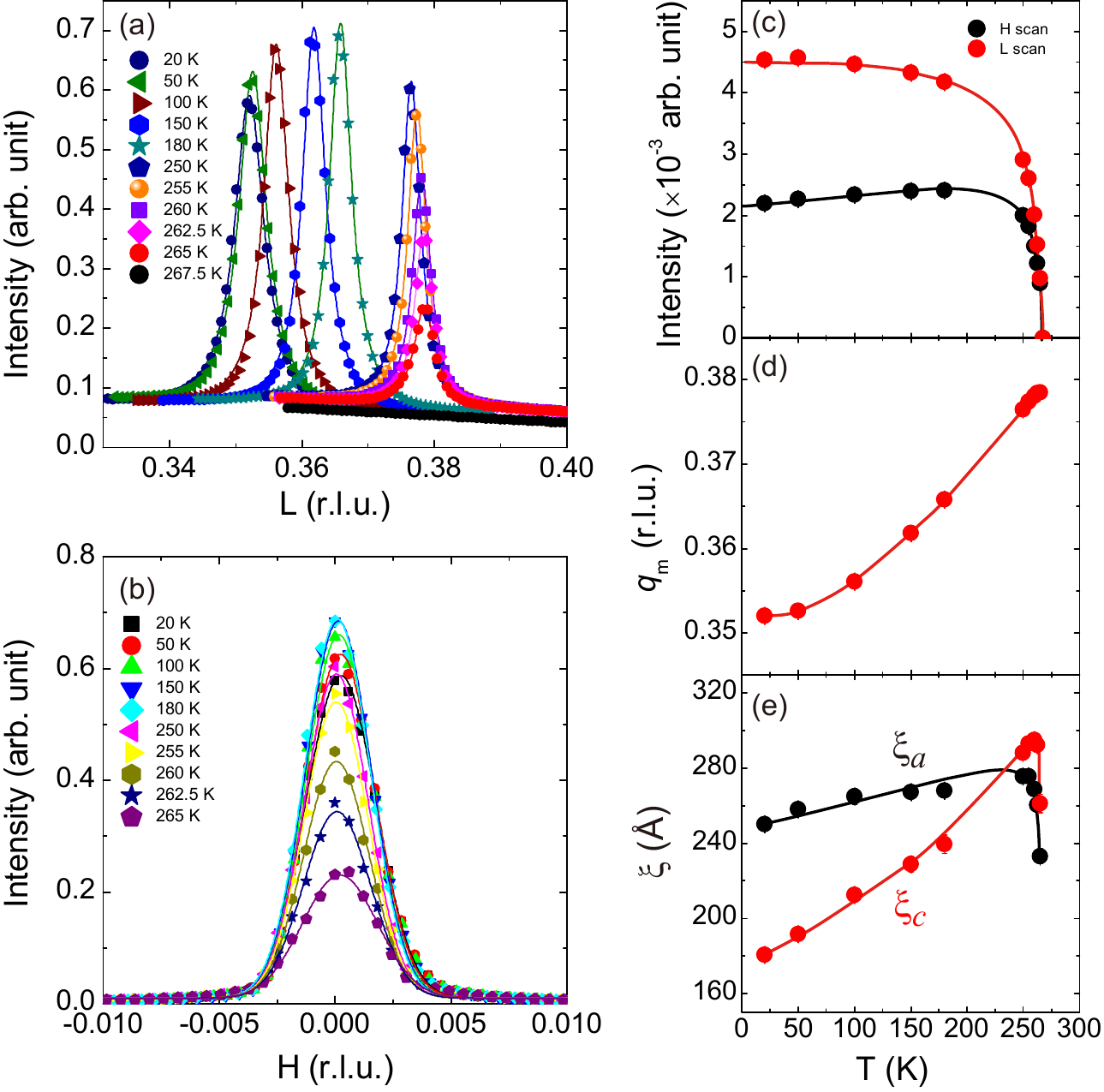}
\caption{\label{fig:epsart} Temperature-dependent evolution of the (0~0~$q_m$) resonant peak along (a) $L$ and (b) $H$ directions in the temperature range from 267.5 K to 20 K. The incident photon energy is 578.7 eV with $\pi$ polarization. The sample was warmed up from 20 K to 267.5 K. The solid lines in (a) and (b) are Lorentzian and Gaussian function fittings, respectively. (c) Integrated intensity of resonant peak, (d) wavevector $q_m$, and (e) the average helimagnetic domain size $\xi$ versus temperature for the $L$ (red dots) and $H$ (black dots) scans. The solid lines are guides to the eye.}
\end{figure}

Detailed study on the temperature-dependent evolution of (0 0 $q_m$) resonant peak was conducted by using the $\pi$-polarized incident photon with the resonant energy 578.7 eV. Both $L$ and $H$ scans were performed at the temperature range from 267.5 K to 20 K, as shown in Fig. 3(a) and 3(b). Along $L$ scans, $q_m$ continuously decreases on cooling, consistent with our neutron diffraction results shown in the Supplemental Material. The $L$ and $H$ scans take the Lorentzian and Gaussian lineshapes, respectively [Fig. 3(a) and 3(b)], indicating differnt domain size distribution along the $c$ and $a$ axes. 
This anisotropy may come from the unique propagating direction of the helimagnetic wavevector or the elongated needle-like crystal shape, both of which are along $c$ axis and could cause anisotropic grain and strain distribution inside the sample.

The temperature-dependent evolutions of peak intensity, propagation wavevector $q_m$, and the average helimagnetic domain size $\xi$ ($\xi$=1/FWHM, FWHM is for full-width-at-half-maximum) in $L$ and $H$ scans are presented in Fig. 3(c)-3(e). In Fig. 3(c), the magnetic peak intensity rapidly saturates below $T_S$, consistent with the first order transition character\cite{Shen2016}. The change of $q_m$ with decreasing temperature (Fig. 3(d)) indicates that the balance of competition between different magnetic interactions in CrAs varies with temperature.
The average helimagnetic domain size $\xi_c$ and $\xi_a$ rapidly grow from $T_S$ to 255 K (Fig. 3(e)), consistent with the typical critical behavior near a transition point\cite{PhysRevB.87.134407}. However, below $\sim$255 K both $\xi_c$ and $\xi_a$ abnormally $decrease$ on cooling (Fig. 3(e)). The broadening of the magnetic peak also leads to a slight decreasing of peak intensity along $H$ (Fig. 3(c)). Usually, the average helimagnetic domain size of a magnetic order should monotonically $increase$ below $T_S$ because thermal fluctuations are weakened and spins become more correlated on cooling. The anomalous $\xi$ versus $T$ in CrAs indicates the average helimagnetic domain size actually shrinks with lowering temperature. Cooling and warming sequences (see Supplemental Material) show little thermal history effect in the temperature-dependent evolution of $\xi$ and $q_m$.

The anomalous shrinkage of magnetic domains on cooling can be interpreted by the weakening of DMI in CrAs. In ref. 13, the authors gave a detailed description on the magnetic interactions in CrAs. Its magnetic Hamitonian was represented by Eq. 1, in which $\vec{D}_{i,j}$ and $J_{i,j}$ are the DMI and antiferromagnetic interactions between the nearest neighbors, respectively. The nearest-neighboring spins in a single unit cell are illustrated by the red dashed lines in Fig. 1(b). $q_m$ can be expressed as\cite{Shen2016}:
\begin{equation}
q_m=\frac{\beta_{12}+\beta_{23}}{\pi}
\end{equation}
where $\beta_{12}$ is the angle between $\vec{S}_1$ and $\vec{S}_2$, $\beta_{23}$ is the angle between $\vec{S}_2$ and $\vec{S}_3$. In the temperature range of our study, $\beta_{23}$ barely changes \cite{Shen2016}. Therefore, the decrease of $q_m$ on cooling is mainly attributed to the variation of $\beta_{12}$, which is determined by:
\begin{equation}
\beta_{12}=\mathrm{tg}^{-1}(D_{12}^c/J_{12})
\end{equation}
in which $D_{12}^c$ is the $\vec{D}_{12}$ component along the $c$ axis. There are antiferromagnetic interactions between all nearest spins, in contrast, DMI exists between $\vec{S}_1$ and $\vec{S}_2$ but is absent between $\vec{S}_2$ and $\vec{S}_3$\cite{Shen2016}. Moreover, the DMI in CrAs is exceptionally larger than the antiferromagnetic interaction ($\mid$$D$$\mid$$>$$\mid$$J$$\mid$) \cite{Kallel1974}. Therefore, the induced change of $\beta_{12}$ and the decreasing of $q_m$ should be dominated by the variation of $D_{12}^c$:
\begin{equation}
\Delta(q_m)\propto\Delta(D_{12}^c)
\end{equation}
According to this equation, the 6.70\% decrease of propagation wavevector, from $q_m$=0.3773(5) at 255 K to $q_m$=0.3520(6) at 20 K, indicates that $D_{12}^c$ become weaker. Since $D_{12}^c$ favors non-collinear spin alignment and $J_{12}$ favors antiparallel spin alignment in CrAs \cite{Shen2016}, $\vec{S}_1$ and $\vec{S}_2$ tends to be more antiparallel\cite{Shen2016}, as illustrated by $\beta_{12}$ in Fig. 4. This again evidences the weakening of DMI with decreasing temperature. As DMI is the dominant force determining the spin rotation along the helix chain, its weakening will make the helimagnetic domains easier to break up at the defect sites. As $q_m$ varies with temperature, the neighboring spins continuously modulate their relative spin angles on cooling, which would generate additional domain boundaries at defect sites given the weakening DMI, in other words, the helimagnetic domains shrink. A straightforward cartoon illustration for the DMI controlled spin angle $\beta_{12}$ and the accompanied domain shrinkage is presented in Fig. 4.

\begin{figure}[]
\centering
\includegraphics[clip,width=8.5cm]{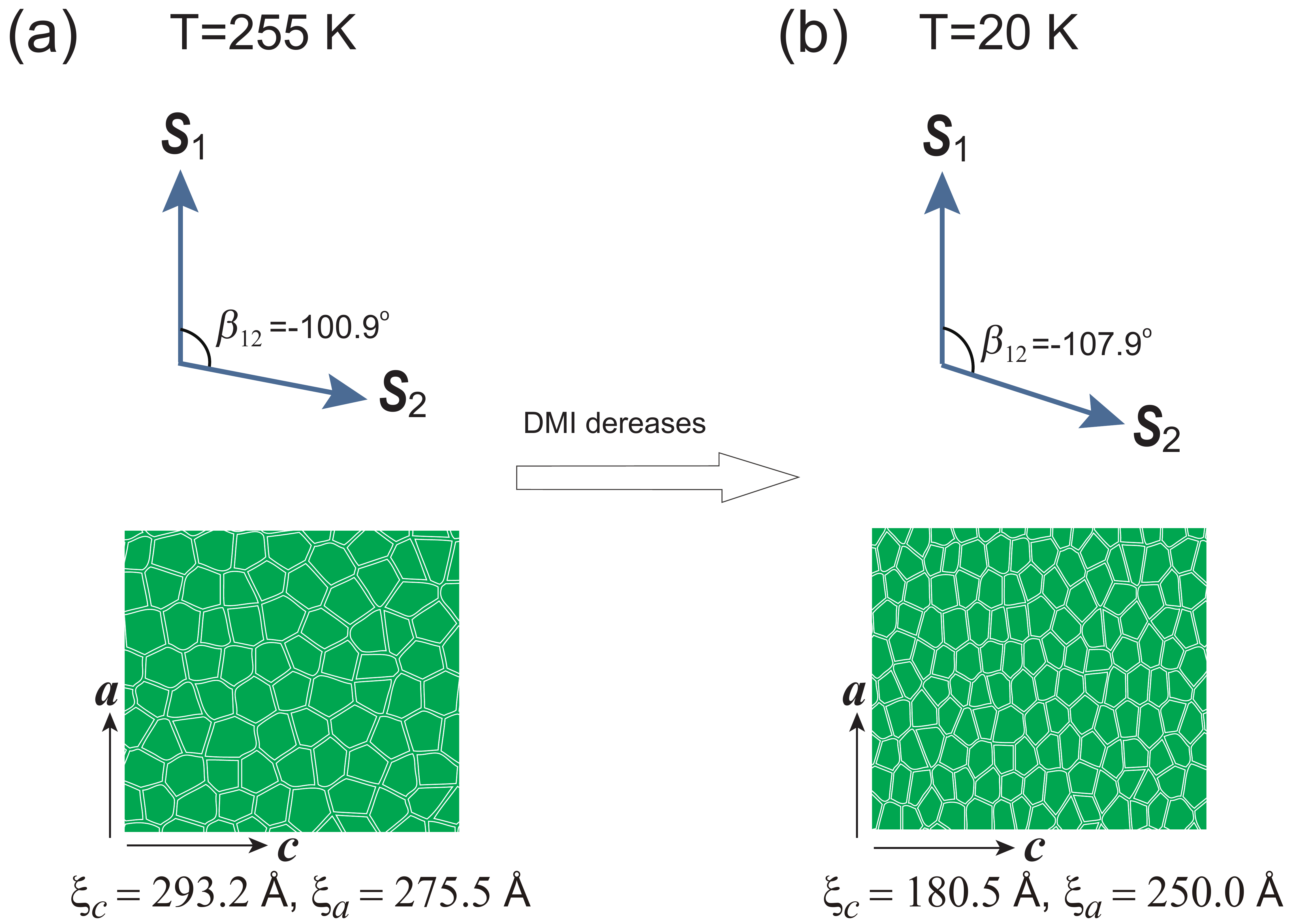}
\caption{\label{fig:epsart} Cartoon illustration of the angle ($\beta_{12}$) between $\vec{S}_1$ and $\vec{S}_2$ and helimagnetic domains in the $ac$ plane at (a) 255 K and (b) 20 K. The values of $\beta_{12}$ are from ref. 13. As the DMI term in the magnetic Hamiltonian has the form $\vec{D}_{12}$$\cdot$($\vec{S}_1\times\vec{S}_2$), thus $\vec{D}_{12}$ favours non-collinear spin alignment. The tendency to antiferromagnetic spin alignment between $\vec{S}_1$ and $\vec{S}_2$ from 255 K to 20 K indicates the DMI gets weaker on cooling.}
\end{figure}

The helimagnetic domain shrinkage is anisotropic and mainly takes place along the $c$ direction. As shown in Fig. 3(e), from 255 K to 20 K the percentage drop of $\xi_c$ and $\xi_a$  are 38.44 \% and 9.26 \%, respectively. Here we define the spatial anisotropic ratio of domain shrinkage as $\gamma_{ca}$=$\frac{\Delta\xi_c/\xi_c(255 \mathrm{K})}{\Delta\xi_a/\xi_a(255 \mathrm{K})}$=4.15.
This is consistent with the fact that the helimagnetic order is propagating along the \textit{c} direction. Meanwhile, it is intriguing to note that
the DMI of CrAs is $\vec{D}_{12}\approx D_0$(-0.17, -0.5, 0.85)\cite{Shen2016}, so the ratio of DMI components along $c$ and $a$ is $\kappa_{ca}=\vert D_{12}^c/D_{12}^a\vert$=5. The similar size of $\gamma_{ca}$ and $\kappa_{ca}$ implies possible role played by DMI in the anisotropy of domain shrinkage, as certain interactions exist in \textit{a} and \textit{b} directions as well.

By contrast, in our previous RSXS investigation on MnP, a helimagnet similar to CrAs in lattice and magnetic structures but its propagation wavevector increases on cooling\cite{Motizuki2010}, the domain shrinkage behavior was not observed\cite{Pan2018}. This implies that the decrease of $q_m$ or DMI on cooling is the key to the formation of new domain boundaries at defect sites inside the sample. It should be noted that in most 3$d$-transition metal pnictides the strength of DMI is much smaller than $J$, however, CrAs is an exception in which $\mid$$D$$\mid$$>$$\mid$$J$$\mid$ \cite{Kallel1974}. The strong DMI of CrAs even drives the spin reorientation transition and decrease of magnetic wavevector under pressure\cite{Shen2016}.
Therefore, we conclude that the pronounced DMI strength combined with its decrease on cooling are essential ingredients for the anomalous helimagnetic domain shrinkage behavior in CrAs.
Broadening of magnetic peak at low temperature was also observed in Ca$_3$Co$_2$O$_6$ with ferromagnetic chains, while it is attributed to the development of a short range order \cite{Agrestini2008}, which is distinct from the single-component magnetic peak in CrAs and does not involve DMI.

In summary, we find the Cr$^{3+}$ valance state in CrAs and our RSXS experiment reveals its helimagnetic domains shrink on cooling below $\sim$255 K. The domain shrinkage has similar temperature-dependent evolution with that of DMI, indicating DMI is the main driving force in this anomalous behavior. Our results reveal a quantum effect that is opposite to conventional thermal effect, and suggest that DMI may be tuned to manipulate the domain boundaries in helimagnets which may have application in future spintronics.

The authors are grateful for the helpful discussions with Prof. Jiang Xiao, Prof. Yi-Zheng Wu, and Prof. Yan Chen of Fudan University. This work is supported by the National Natural Science Foundation of China (Grant Nos.\ 11888101, 11790312, 11804137, 11704074), the National Key Research and Development Program of China (Grant No.\ 2016YFA0300200 and No.\ 2017YFA0303104), the Science and Technology Commission of Shanghai Municipality (Grant No.\ 15ZR1402900), and the Natural Science Foundation of Shandong Province (Grant No.\ ZR2018BA026). Part of the research described in this paper was performed at the Canadian Light Source, a national research facility of University of Saskatchewan, which is supported by the Canada Foundation for Innovation (CFI), the Natural Sciences and Engineering Research Council (NSERC), the National Research Council (NRC), the Canadian Institutes of Health Research (CIHR), the Government of Saskatchewan, and the University of Saskatchewan.

\bibliography{CrAs}

\clearpage
\begin{figure}[]
\centering
\includegraphics[clip,width=19cm]{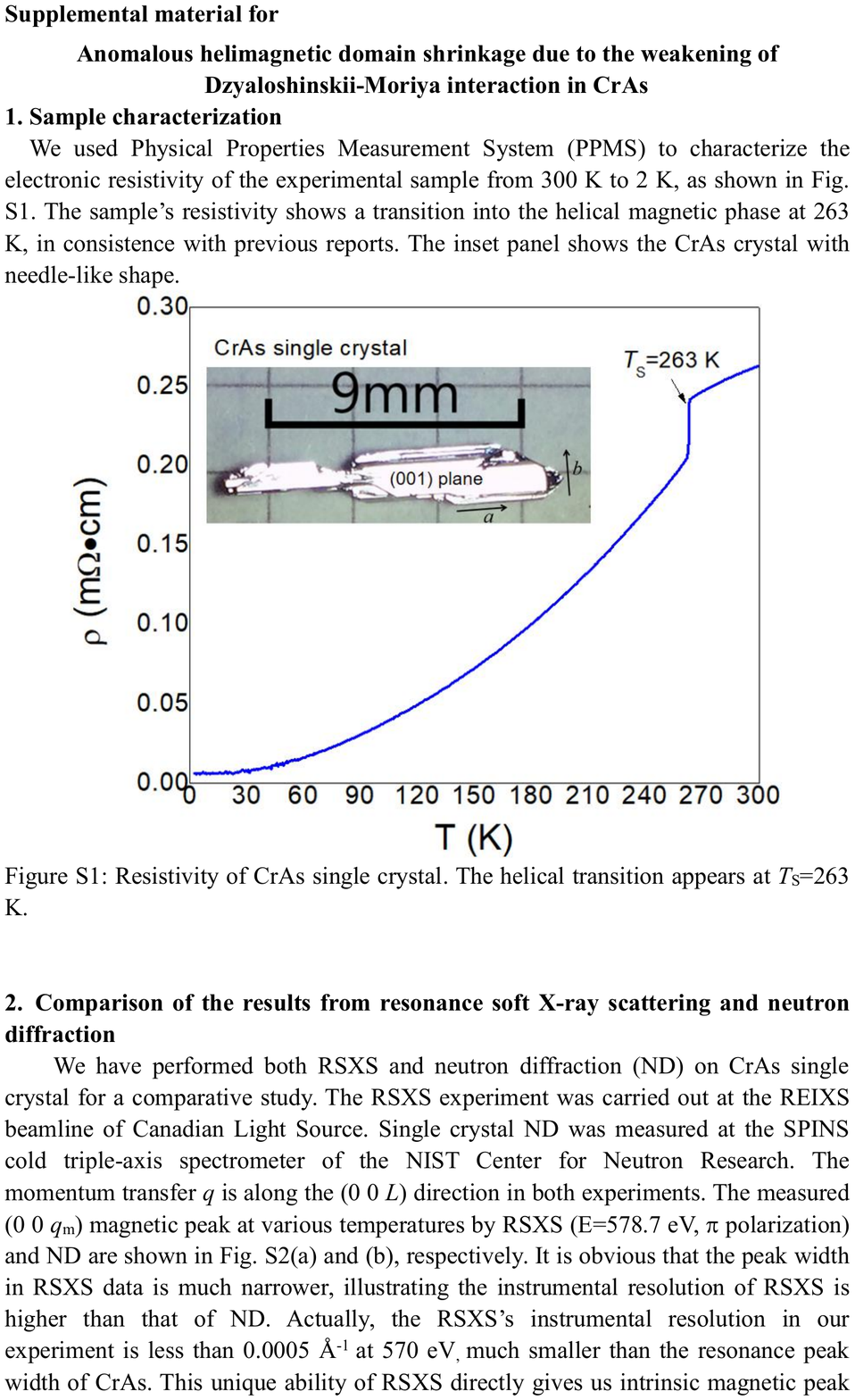}
\caption{}
\end{figure}

\clearpage
\begin{figure}[]
\centering
\includegraphics[clip,width=19cm]{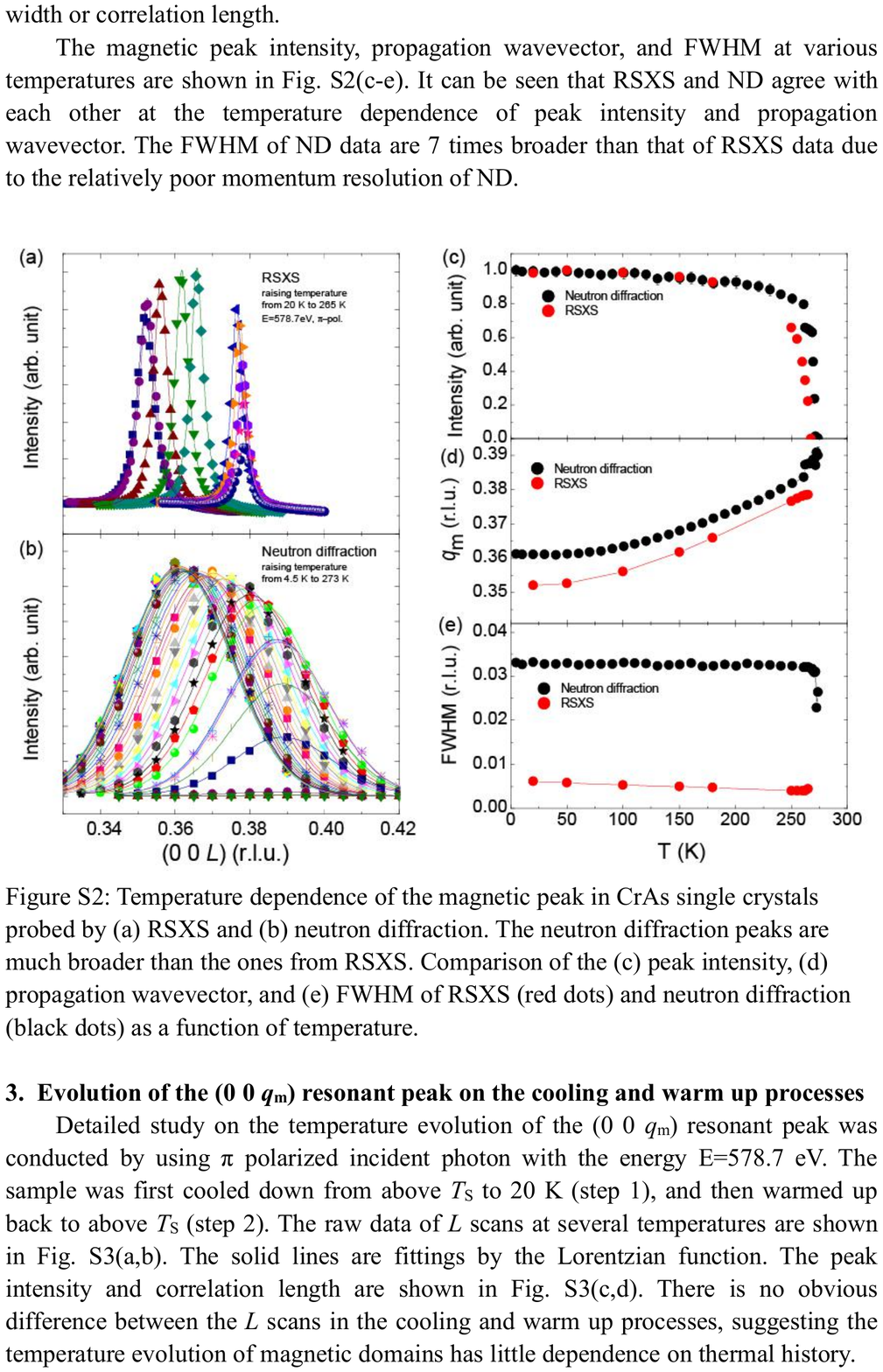}
\caption{}
\end{figure}

\clearpage
\begin{figure}[]
\centering
\includegraphics[clip,width=19cm]{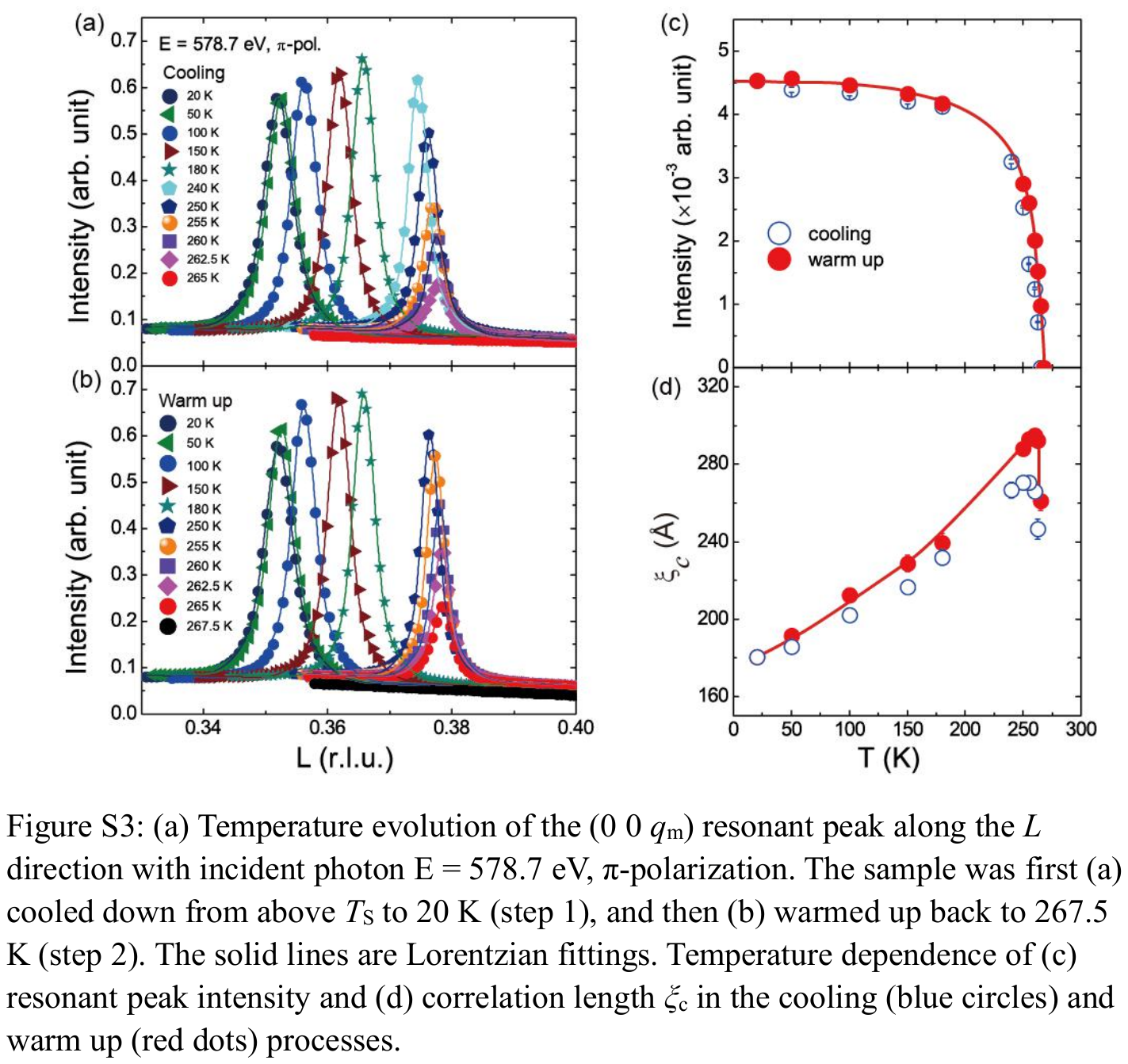}
\caption{}
\end{figure}
\end{document}